\documentclass[aps,pra,reprint,groupedaddress]{revtex4-1}
\usepackage{graphicx}
\usepackage{color}
\usepackage{bm}
\usepackage{comment}

\begin{document}


\title{Reservoir Computing on Spin-Torque Oscillator Array}


\author{Taro Kanao}
\email[]{taro.kanao@toshiba.co.jp}
\author{Hirofumi Suto}
\author{Koichi Mizushima}
\author{Hayato Goto}
\author{Tetsufumi Tanamoto}
\author{Tazumi Nagasawa}
\affiliation{Corporate Research and Development Center, Toshiba Corporation, 1, Komukai-Toshiba-cho, Saiwai-ku, Kawasaki 212-8582, Japan}


\date{\today}

\begin{abstract}
We numerically study reservoir computing on a spin-torque oscillator (STO) array, describing the magnetization dynamics of the STO array by a nonlinear oscillator model.
The STOs exhibit synchronized oscillation due to coupling by magnetic dipolar fields.
We show that reservoir computing can be performed using the synchronized oscillation state.
The performance can be improved by increasing the number of STOs.
The performance becomes highest at the boundary between the synchronized and disordered states.
Using an STO array, we can achieve higher
performance than that of an echo-state network with similar number of units.
This result indicates that STO arrays are promising for hardware implementation of reservoir computing.
\end{abstract}


\maketitle

\section{Introduction}\label{sec_intro}
Reservoir computing~\cite{Jaeger2004, Maass2002, Verstraeten2007}, a framework for machine learning, has attracted much attention for realizing high-performance real-time computing for applications such as time-series modeling and prediction, signal processing, and pattern classification.
Reservoir computing utilizes the complex dynamics of interacting nonlinear units.
In reservoir computing, only output weights are learned in an external computer.
This feature allows hardware implementations of reservoir computing, as well as reduction of the computational costs of the learning process.
Hardware implementations have been proposed in several physical systems, such as analog electronic circuits~\cite{Appeltant2011}, optical systems~\cite{Larger2012, Paquot2012}, quantum-bit systems~\cite{Fujii2017}, and magnetic systems~\cite{Prychynenko2018, Nakane2018, Furuta2018a}.
Although these systems have shown good potential for reservoir computing, further improvements in computational capacity, operation speed, hardware size, energy consumption, and balance are required for practical applications.

Recently, reservoir computing on spin-torque oscillators (STOs)~\cite{Kiselev2003, Slavin2009} has been reported~\cite{Torrejon2017a, Tsunegi2018a, Markovic2019, Tsunegi2019}.
(Applications of STOs to other types of computing have also been proposed~\cite{Macia2011, Shibata2012, Csaba2013, Grollier2016a, Kudo2017, Vodenicarevic2017b, Romera2018}.)
An STO is a nonlinear auto-oscillator with a power-dependent frequency shift.
An STO exhibits microwave oscillation induced by a spin-polarized direct current via a spin-transfer torque, yielding microwave magnetic and electromagnetic fields.
Compared with the other systems above, STOs have the advantages of nanosecond-scale fast dynamics, nanometer-scale size, and low energy consumption.
The microwave fields can lead STOs to interact with each other in an array.
Although the use of many interacting STOs is expected to achieve higher performance due to complex dynamics with many decrees of freedom, a {\it single} STO has been assumed in previous studies of STO reservoir computing~\cite{Torrejon2017a, Tsunegi2018a, Markovic2019, Tsunegi2019}. 

In this paper, we investigate reservoir computing on an array of STOs coupled by magnetic dipolar fields, as shown in Fig.~\ref{fig_sto_array}, where couplings naturally appear when the STOs are integrated closely with each other.
We perform numerical experiments on reservoir computing utilizing a nonlinear oscillator model~\cite{Slavin2009} corresponding to an STO array.
As a result, we show that reservoir computing can be performed in the synchronized oscillation state of the array and that its performance can be improved by increasing the number of STOs.
The highest performance is obtained at the boundary between the synchronized and disordered states.
The performance can become higher than that of a standard neural-network model known as an echo-state network (ESN)~\cite{Jaeger2004, Maass2002}.

\section{Reservoir computing on STO array}
\subsection{Model of STO array}
\begin{figure}
	\centering
	\includegraphics[width=8cm,bb=0 0 433 290]{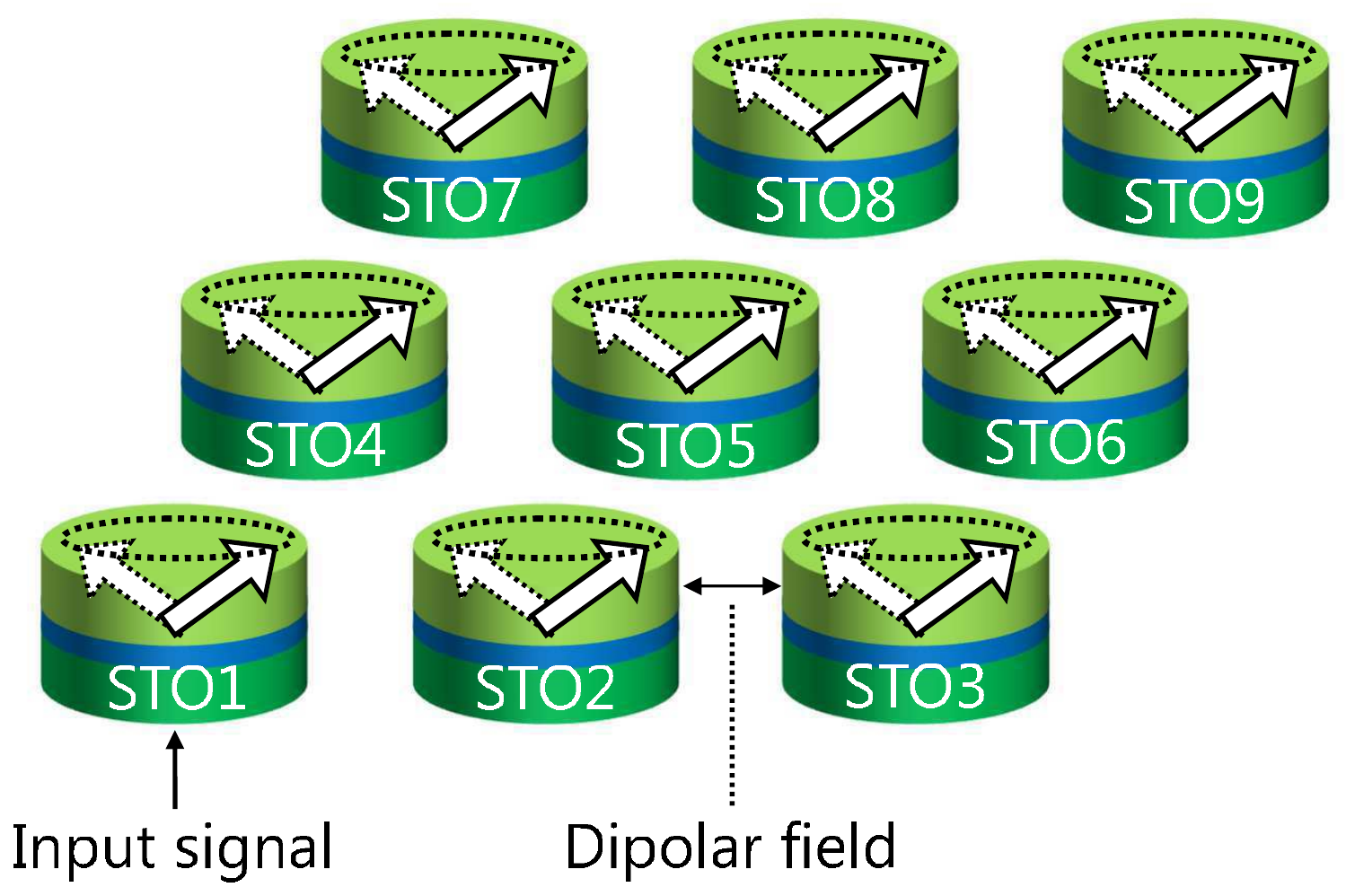}
	\caption{Schematic of a spin-torque oscillator array.}
\label{fig_sto_array}
\end{figure}

We place $N$ STOs on a square lattice with a size of ${M\times M(=N)}$ as shown in Fig.~\ref{fig_sto_array}.
In this paper, we assume that the STOs exhibit out-of-plane precession~\cite{Kiselev2004}.
These STOs emit large microwave dipolar fields~\cite{Kanao2018} and output signals due to the magnetoresistive effect~\cite{Zeng2013a, Kubota2013}, which lead to strong couplings between the STOs and a high signal-to-noise ratio in measurements, respectively.
The oscillations of the STOs are synchronized, with the same phase~\cite{Belanovsky2012, Locatelli2015a, Flovik2016}.
[We show in Appendix~\ref{sec_sto_llgs} an example of such synchronized oscillations by use of a numerical simulation based on the Landau-Lifshitz-Gilbert-Slonczewski (LLGS) equation~\cite{Slonczewski1996}.]

For reservoir computing, we utilize the dynamics of the oscillation powers and phases (or frequencies) of the STOs~\cite{Torrejon2017a, Tsunegi2018a, Markovic2019, Tsunegi2019}.
To calculate this dynamics, we use a corresponding nonlinear oscillator model~\cite{Slavin2009, Kanao2016}, which can capture the essential features of the dynamics of the STOs, including the synchronization~\cite{Slavin2006}, irrespective of the detailed magnetization configuration of the STOs.
The nonlinear oscillator model can be derived from the LLGS equation under certain assumptions.
(See Appendix~\ref{sec_llgs_nlom} for details.)
The powers and frequencies of the STOs usually vary slowly compared with the oscillation itself~\cite{Slavin2009, Kanao2018}, and fast oscillation components can be eliminated in the nonlinear oscillator model.
Thus this model allows us to reduce the computational costs of the simulation.

The nonlinear oscillator model for $N$ STOs is written as~\cite{Slavin2009}
\begin{eqnarray}
	\frac{\mathrm{d}c_i}{\mathrm{d}t}=-\mathrm{i}\delta\omega_i(p_i)c_i-\Gamma_i(p_i)c_i-\mathrm{i}\sum^N_{j=1,j\neq i}\Omega_{ij}c_j,\label{eq_nlo1}
\end{eqnarray}
where ${t, c_i, \delta\omega_i(p_i), \Gamma_i(p_i)}$, and $\Omega_{ij}$ (${i=1,2,\cdots,N}$) are the time, complex amplitudes representing the slowly-varying magnetization dynamics, the nonlinear frequency shifts originating from magnetic anisotropy, the effective damping rates taking account of Gilbert damping and the spin-transfer torque, and the coupling strengths between the STOs caused by the dipolar fields, respectively.
Here, a uniform high-frequency oscillation factor has been eliminated (see Sec.~\ref{sec_coupling}).
$\delta\omega_i(p_i)$ and $\Gamma_i(p_i)$ are supposed to depend on a dimensionless oscillation power ${p_i=\left|c_i\right|^2}$.
Assuming a linear dependence of $\delta\omega_i(p_i)$ and $\Gamma_i(p_i)$ on $p_i$~\cite{Slavin2009, Kanao2016}, we express them as
\begin{eqnarray}
	\delta\omega_i(p_i)&=&N_f\left(p_i-p_{i0}\right),\label{eq_omega}\\
	\Gamma_i(p_i)&=&\frac{\Gamma_p}{p_{i0}}\left(p_i-p_{i0}\right)-\xi_i(t),\label{eq_gamma}
\end{eqnarray}
where $p_{i0}, N_f, \Gamma_p$, and ${\xi_i(t)}$ are the stationary oscillation power, the nonlinear frequency shift, the rate of damping of power, and the modulation by an input signal, respectively, in units of frequency~\cite{Slavin2009}.
$\delta\omega_i(p_i)$ and $\Gamma_i(p_i)$ describe the properties of a single STO, and ${\delta\omega_i(p_{i0})=\Gamma_i(p_{i0})=0}$ holds for its stationary oscillation state for a constant current.
We add the input signal to the current, which, via the spin-transfer torque, leads to a modulation of $\Gamma_i(p_i)$.
This modulation is represented by $-\xi_i(t)$.
It is supposed that $\delta\omega_i(p_i)$ does not directly depend on the current, and instead is affected through the changes in $p_i$~\cite{Slavin2009}.
Equations~(\ref{eq_omega}) and~(\ref{eq_gamma}) can be regarded as linear approximations of $\delta\omega_i(p_i)$ and $\Gamma_i(p_i)$ for small deviations of $p_i$ from $p_{i0}$ and for an input signal in the current.
(See Sec.~\ref{sec_EffectiveDamping} for details.)

In the array, the dipolar fields between the nearest-neighbor STOs $\Omega$, and between the next-nearest-neighbor STOs, ${\Omega/\sqrt{8}}$, are taken into account, as the dipolar field is inversely proportional to the cube of the distance.
Differences between the STOs are modeled by taking $p_{i0}$ as a Gaussian random variable with a mean of $p_0$.
In this work, the input signal is applied to an STO at the corner of the array, namely STO1 in Fig.~\ref{fig_sto_array}, and thus only $\xi_1(t)$ is modulated, and ${\xi_i(t)=0}$ for ${i\neq1}$.
Based on reported values~\cite{Kanao2016, Quinsat2010}, we choose the following values: ${p_0=0.5}$, ${N_f/(2\pi)=0.1}$ GHz, ${\Gamma_p=0.1}$ ns$^{-1}$, ${\left|\xi_1(t)\right|<0.1}$ ns$^{-1}$, and ${\Omega=-0.1}$ ns$^{-1}$, unless otherwise stated.
The value of $\Omega$ is estimated in Sec.~\ref{sec_coupling}.
The standard deviation of $p_{i0}$ is set to 0.01.
The denominator $p_{i0}$ in Eq.~(\ref{eq_gamma}) is approximated by $p_0$.
Equation~(\ref{eq_nlo1}) is numerically integrated with the initial condition ${c_i=\sqrt{p_0}}$, which is equivalent to all the magnetizations of a free layer of STOs being in the same direction.
This situation can be realized in a strong uniform external field.

\begin{figure}[b]
	\centering
	\includegraphics[width=8.5cm,bb=0 0 377 296]{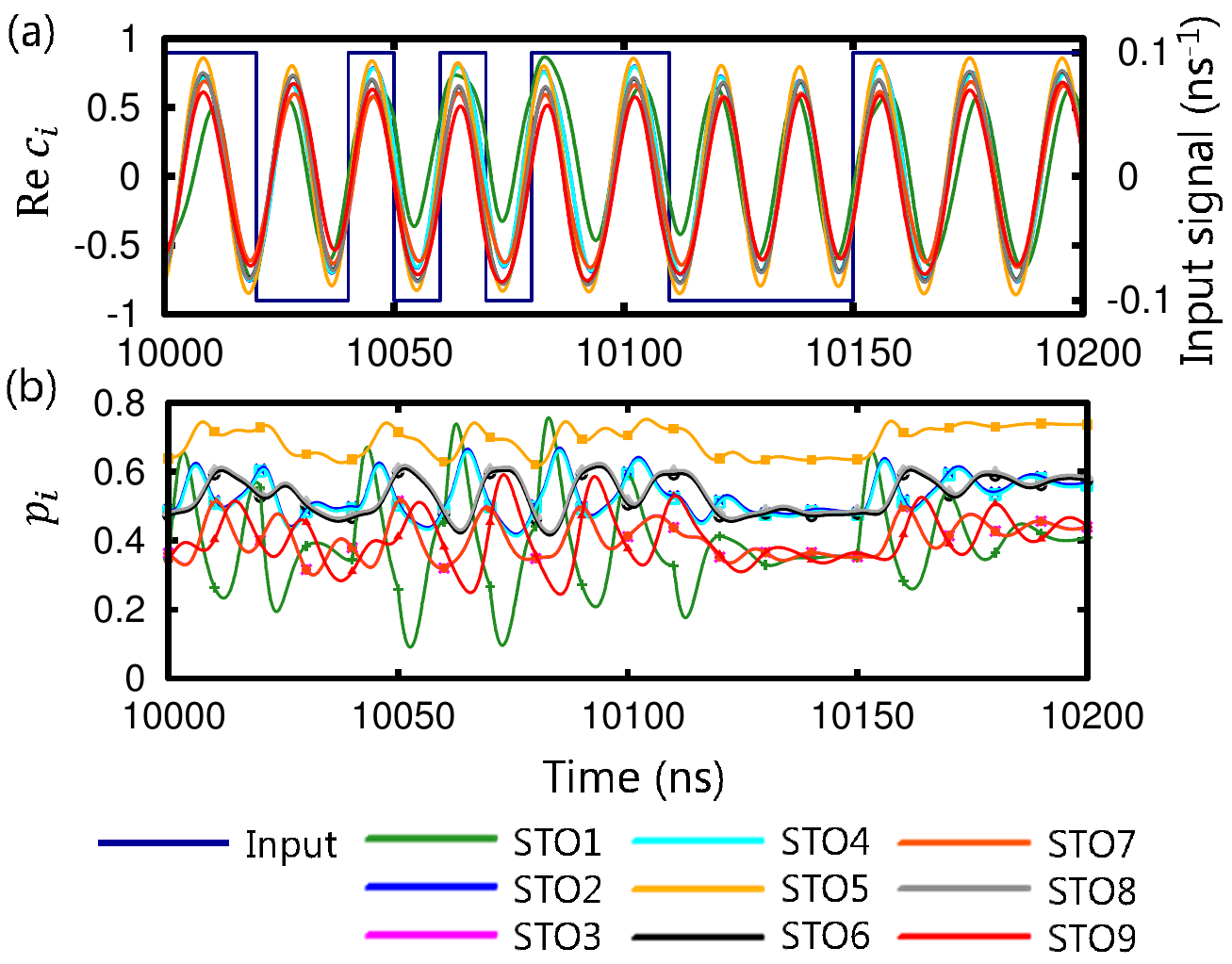}
	\caption{(a) Waveforms of input signal $\xi_1(t)$ and real parts of $c_i$ for nine STOs with a sampling time ${T_s=10}$ ns.
	(b) Oscillation powers $p_i$ under the same conditions.
		Measured values of $p_i$ are indicated by symbols.}
\label{fig_unita_pow_time}
\end{figure}
Reservoir computing utilizes the dynamics of the waveforms induced by the input signals.
Figure \ref{fig_unita_pow_time} shows examples of waveforms of the input signal $\xi_1(t)$ and of nine STOs obtained from Eq.~(\ref{eq_nlo1}).
The waveforms after a sufficiently long time are shown.
The oscillation of $\mathrm{Re}\hspace{0.2em}c_i$ in Fig.~\ref{fig_unita_pow_time}(a) is induced by the terms $\delta\omega_i(p_i)$ and $\Omega_{ij}$ in Eq.~(\ref{eq_nlo1}).
It can be seen that the oscillation phases are nearly synchronized.
Fluctuations of the oscillations are induced by the input signal.
Since the input signal is applied to STO1, the power of STO1 is directly modulated by the input signal, as shown in Fig.~\ref{fig_unita_pow_time}(b).
The effects of the input signal propagate to the other STOs.
When the input signal is constant, the dynamics of the powers is damped in several tens of nanoseconds.
The timescale of the damping of the powers is determined by ${1/\Gamma_p(=10}$ ns).
The waveforms of the STOs at symmetric positions (see Fig.~\ref{fig_sto_array}), i.e., (STO2 and 4), (STO3 and 7), and (STO6 and 8), almost overlap, but not completely, because of the deviation of $p_{i0}$.

\subsection{Reservoir computing}
Next, we explain reservoir computing on the STO array.
We evaluate its performance by following the setting used in Ref.~\cite{Fujii2017}.
The $k$th input signal, $s_k$, is introduced into STO1 at a rate $1/T_s$ during the time interval between ${T_s(k-1)}$ and $T_sk$.
The powers are measured at the end of each interval, in order to fully incorporate the effects of the latest signal.
The measured values are denoted by ${x_{ki}=p_i(T_sk)}$.
The output data are ${y_k=\sum^{N}_{i=0}x_{ki} w_i}$, where $w_i$ is the $i$th weight for reservoir computing, and ${x_{k0}=1}$ is introduced to treat a constant term.
The weights are determined in a training process by minimizing $\sum_k(y_k-\bar{y}_k)^2$, so that a target series $\bar{y}_k$ is reproduced.
These $w_i$ are given by ${{\bf w}=X^+ \bar{{\bf y}}}$, where $X$ is a matrix and ${\bf w}$ and $\bar{{\bf y}}$ are column vectors, with components $x_{ki}$, $w_i$, and $\bar{y}_k$, respectively, and where $X^+$ is the Moore-Penrose pseudoinverse matrix of $X$~\cite{Fujii2017}.
The dynamics of the STO array is calculated for a time of $5000T_s$.
The first $1000T_s$ is discarded, and the subsequent $3000T_s$ and the last $1000T_s$ are used for training and evaluation, respectively.

We evaluate the performance by using short-term memory (STM) and parity-check (PC) tasks~\cite{Fujii2017, Furuta2018a, Tsunegi2018a, Tsunegi2019}, because these tasks can estimate capacities for short-term memory and nonlinearity, which are necessary for applications to real-time computing.
The STM task measures the STM capacity by outputting delayed inputs.
The PC task can estimate nonlinearity because the required function is a nonlinear mapping.
[We also evaluate a nonlinear autoregressive moving average (NARMA) task~\cite{Fujii2017,Atiya2000} and show the result in Appendix~\ref{sec_narma}.]
For the STM and PC tasks, $s_k$ is a random sequence of $0$s and $1$s.
We convert $s_k$ to $\xi_1(t)$ by use of a square wave with value ${\xi_1(t)=0.1(2s_k-1)}$ ns$^{-1}$.
The targets of the STM and PC tasks are ${\bar{y}_k=s_{k-\tau}}$ and ${\bar{y}_k=Q\left(\sum^\tau_{l=0}s_{k-l}\right)}$, respectively, for ${\tau=0,1,2,\cdots}$.
The parity function $Q(S)$ is 0 for even $S$ and 1 for odd $S$.
The capacities are measured by ${C=\sum^{\tau_{\rm max}-1}_{\tau=0}C_\tau}$, where $C_\tau$ is the square of the correlation coefficient of $\bar{y}_k$ and $y_k$ called a $\tau$-delay capacity ${\left(0\leq C_\tau\leq1\right)}$.
Here, ${\tau_{\rm max}=500}$.
A larger $C_\tau$ means higher performance, and $C_\tau$ reaches its maximum value of 1 when ${\bar{y}_k=y_k}$.
The performances are evaluated for 20 sets of $\left(p_{i0},s_k\right)$, and the average values of the performances and standard deviations are calculated.

\section{Results and discussion}
\subsection{Performance improvement with increasing $N$}
\begin{figure}
	\centering
	\includegraphics[width=8.5cm,bb=0 0 243 460]{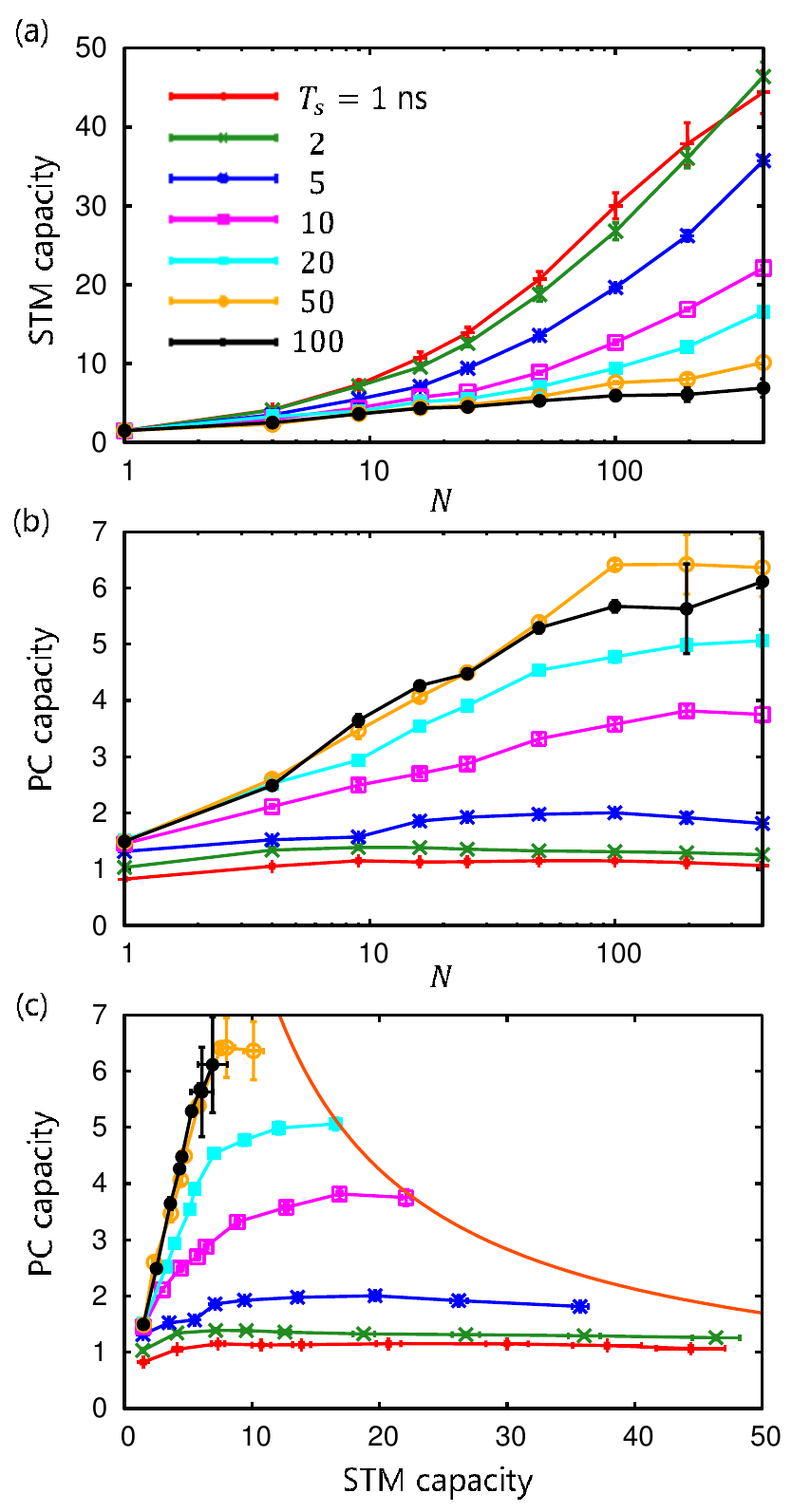}
	\caption{(a) Short-term memory and (b) parity-check capacities as a function of number of STOs $N$ and sampling time $T_s$.
		$T_s$ is indicated by common symbols and colors in (a)-(c).
		(c) STM capacity in (a) vs PC capacity in (b).}
\label{fig_stm_pc_n}
\end{figure}

Figures~\ref{fig_stm_pc_n}(a) and \ref{fig_stm_pc_n}(b) show the STM and PC capacities as a function of the number of STOs $N$ and the sampling time $T_s$.
The STM capacity increases with $N$ in the range shown (${N\leq400}$).
The PC capacity saturates at large $N$.
The behaviors of the capacities at larger $N$ are discussed below.
The STM and PC capacities depend on $T_s$ in opposite ways; namely, the STM capacity decreases with increasing $T_s$, while the PC capacity increases.
The STM capacity decreases with increasing $T_s$ because the effect of the input signal decays when $T_s$ is long.
On the other hand, the PC capacity increases with increasing $T_s$ because a longer $T_s$ allows larger variations of the powers such that nonlinear effects appear.
These behaviors occur in the range of $2$ ns ${\lesssim T_s\lesssim50}$ ns, which can be understood from the timescale of the damping of power, ${1/\Gamma_p=10}$ ns.
Outside this range of $T_s$, the powers are no longer susceptible to influence from the input signals.

The data in Figs.~\ref{fig_stm_pc_n}(a) and \ref{fig_stm_pc_n}(b) are summarized in Fig.~\ref{fig_stm_pc_n}(c).
The capacities are distributed below a curve on which the product of the STM and PC capacities is a constant (equal to $85$), implying a trade-off relation between the STM and PC capacities.
This kind of relation has been reported as a memory-nonlinearity trade-off~\cite{Dambre2012}.

\begin{figure}
	\centering
	\includegraphics[width=8.5cm,bb=0 0 237 301]{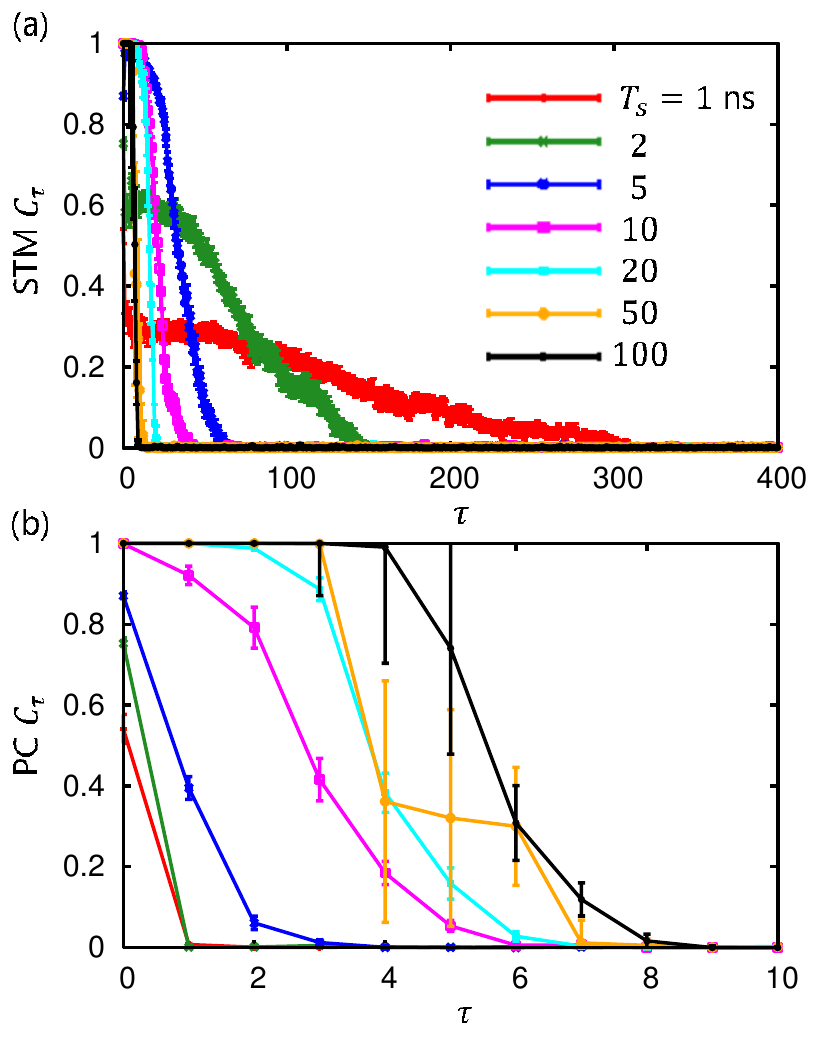}
	\caption{(a) STM and (b) PC capacities $C_\tau$ as a function of $\tau$ for ${N=400}$.}
\label{fig_stm_pc_tau}
\end{figure}
Figure \ref{fig_stm_pc_tau} shows the capacities at each $\tau$ (the $\tau$-delay capacities) for ${N=400}$.
The capacities decrease from 1 to 0 with increasing $\tau$ for ${T_s\geq10}$ ns ${=1/\Gamma_p}$.
The capacities for ${T_s<1/\Gamma_p}$ are small compared with 1 even at small $\tau$, which indicates that $T_s$ is too short for the powers to fully respond to the input signals.
The STM capacities for ${T_s\leq2}$ ns show finite values at large $\tau$, indicating that the effects of the input signals remain.

The largest values of the STM and PC capacities obtained in the present study are $46.4$ at ${N=400}$ and ${T_s=2}$ ns, and $6.4$ at ${N=196}$ and ${T_s=50}$ ns, respectively.
These values are larger than the values for an ESN with 500 units reported in Ref.~\cite{Fujii2017}, which are about $17.6$ and $6.0$, respectively.
Remarkably, the largest value of the STM capacity here is larger than twice the value for the ESN.

A possible reason for this large STM capacity is that the STOs have a phase degree of freedom, where the phases are time integrations of the oscillation frequencies in the past.
In the STO array, the phases affect the time evolution of the powers, and a memory for longer time can be obtained from the powers.
In an ESN, on the other hand, each unit has only one variable and shows damped motion in reservoir computing, which suppresses the past memory.
Studies of reservoir computing on general nonlinear oscillator networks are an important topic for future work.
Another difference which can lead to higher performance of an STO array than of an ESN is in the network structure.
The STOs here are coupled by short-range interaction.
This kind of weak connection allows various cooperative dynamics suitable for reservoir computing~\cite{Jaeger2004}, which is discussed in the next section (Sec.~\ref{sec_states}).
On the other hand, the ESN used for comparison is coupled by all-to-all connections, which tend to suppress the diversity of the dynamics.

\subsection{Performance enhancement near transition}\label{sec_states}
\begin{figure}
	\centering
	\includegraphics[width=8.5cm,bb=0 0 260 530]{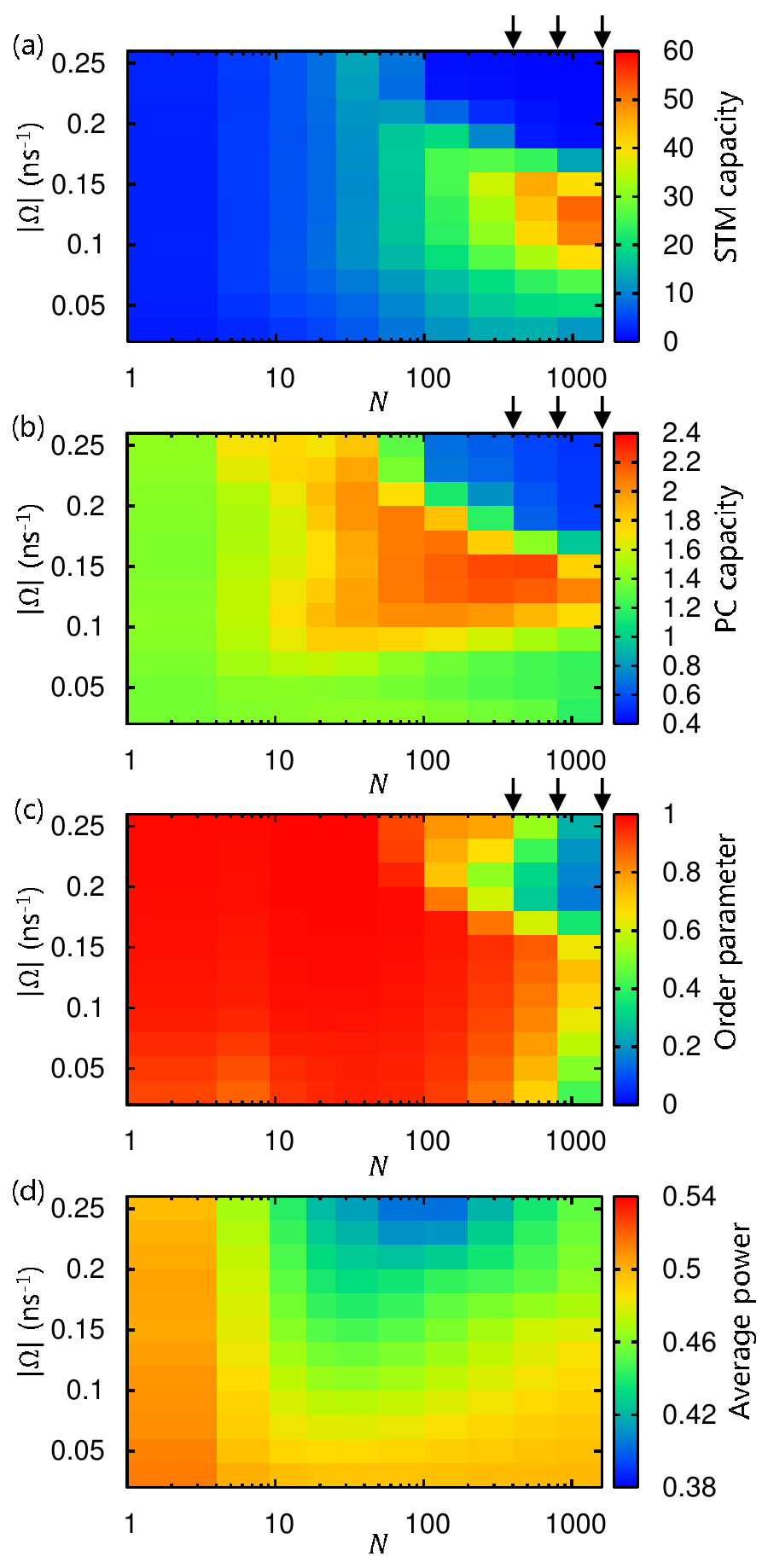}
	\caption{(a) STM and (b) PC capacities, and (c) order parameter $\rho$ and (d) average power ${\bar p}$ as a function of $N$ and coupling strength $|\Omega|$ for ${T_s=5}$ ns.
		Arrows indicate the values of $N$ for which data are plotted in Fig.~\ref{fig_stm_pc_ord_cpl_n}.
		The order parameter and the average power are averaged in the same way as for the STM and PC tasks.
		}
\label{fig_stm_pc_n_cpl}
\end{figure}

Finally, we investigate the dependence of the STM and PC capacities on the coupling, and find that the capacities depend strongly on the coupling strength.
Figures~\ref{fig_stm_pc_n_cpl}(a) and \ref{fig_stm_pc_n_cpl}(b) show the STM and PC capacities as a function of the coupling strength $|\Omega|$ and of $N$ for ${T_s=5}$ ns.
The largest capacities appear around ${|\Omega|=0.14}$ ns$^{-1}$.

To clarify the reason for these dependences of the capacities, we examine the oscillation states of the STO array under the corresponding conditions.
The oscillation states are characterized by an order parameter $\rho$~\cite{Kuramoto1987, Acebron2005} and an average power ${\bar p}$,
\begin{eqnarray}
	\rho&=&\frac{1}{N}\left|\sum^N_{i=1}\frac{c_i}{|c_i|}\right|,\\
	\bar p&=&\frac{1}{N}\sum^N_{i=1}|c_i|^2,
\end{eqnarray}
which are shown in Figs.~\ref{fig_stm_pc_n_cpl}(c) and \ref{fig_stm_pc_n_cpl}(d), respectively.
The order parameter $\rho$ characterizes the uniformity of the oscillation phase, and reaches its largest value of $1$ when all the STOs are perfectly synchronized.

Figure~\ref{fig_stm_pc_n_cpl}(c) shows that $\rho$ is nearly $1$ in most of the region, indicating that all $c_i$ have nearly the same phases and thus the oscillations are synchronized.
$\rho$ decreases as $N$ is increased for the same $|\Omega|$.
The regions where the STM and PC capacities are finite in Figs.~\ref{fig_stm_pc_n_cpl}(a) and \ref{fig_stm_pc_n_cpl}(b) are included in the region where $\rho$ is rather large, ${\left(\rho\gtrsim0.5\right)}$, in Fig.~\ref{fig_stm_pc_n_cpl}(c).

\begin{figure}[b]
	\centering
	\includegraphics[width=8.5cm,bb=0 0 237 475]{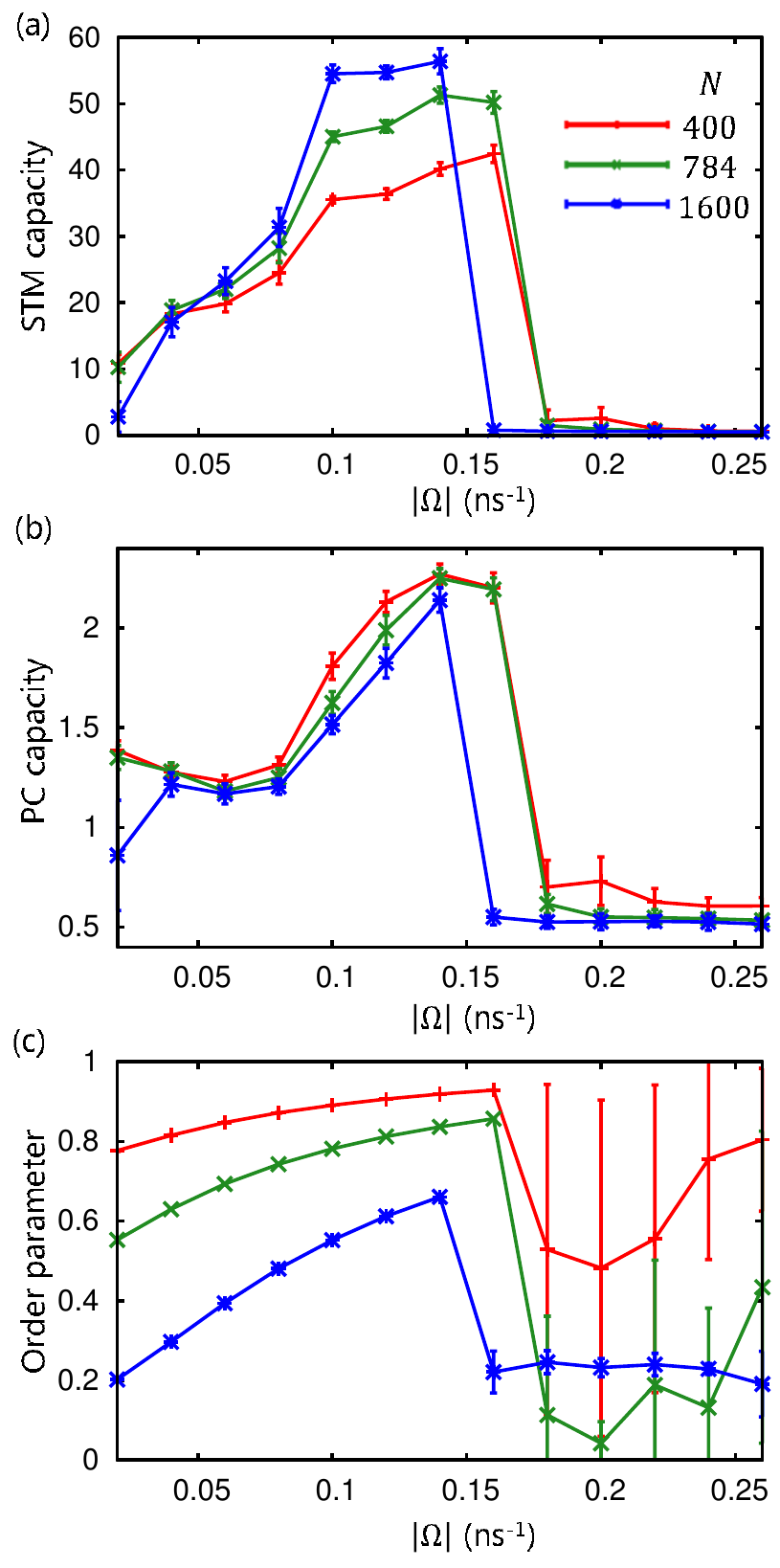}
	\caption{(a) STM and (b) PC capacities, and (c) order parameter $\rho$ as a function of $|\Omega|$ for ${T_s=5}$ ns.
		Error bars indicate standard deviations for 20 sets of $\left(p_{i0}, s_k\right)$.}
\label{fig_stm_pc_ord_cpl_n}
\end{figure}
To see more clearly the relation between the performance and the oscillation state, we plot the capacities and the order parameter as a function of $|\Omega|$ in Fig.~\ref{fig_stm_pc_ord_cpl_n}, where ${N=400(=20^2), 784(=28^2)}$, and ${1600(=40^2)}$.
As shown in Figs.~\ref{fig_stm_pc_ord_cpl_n}(a) and \ref{fig_stm_pc_ord_cpl_n}(b), with increasing $|\Omega|$ the STM and PC capacities increase almost monotonically for ${|\Omega|\lesssim0.15}$ ns$^{-1}$ and then suddenly decrease.
As shown in Fig.~\ref{fig_stm_pc_ord_cpl_n}(c), the order parameter $\rho$ exhibits similar behavior, which indicates a transition from the synchronized oscillation state to a disordered state around ${|\Omega|=0.15}$ ns$^{-1}$ (see below for details).
From the results in Figs.~\ref{fig_stm_pc_n_cpl} and~\ref{fig_stm_pc_ord_cpl_n}, we conclude that reservoir computing can be performed in the synchronized oscillation state of the STO array and the performance is highest near the transition to the disordered state, which is consistent with the reported behavior that the capacities of reservoir computing can be enhanced at the edge of this kind of transition~\cite{Verstraeten2007, Appeltant2011}.

The dependence of the oscillation state on the coupling strength observed in Fig.~\ref{fig_stm_pc_ord_cpl_n}(c) can be understood as follows.
The explanation is divided into two parts, for small and large $|\Omega|$.
For small $|\Omega|$, the power is almost constant, ${\bar p\simeq p_0=0.5}$, as seen in Fig.~\ref{fig_stm_pc_n_cpl}(d), and thus the phase freedom is important.
Then, the model of Eq.~(\ref{eq_nlo1}) can be reduced to a two-dimensional Kuramoto model with a short-range interaction~\cite{Slavin2009, Kuramoto1987, Acebron2005}.
In this Kuramoto model, it is known that stronger couplings lead to more ordered synchronization, while a larger number of oscillators causes a more disordered oscillation state~\cite{Lee2010, Flovik2016}.
These results for the Kuramoto model can explain the dependence of $\rho$ on $|\Omega|$ and $N$ for ${|\Omega|\lesssim0.15}$ ns$^{-1}$ in Fig.~\ref{fig_stm_pc_ord_cpl_n}(c), where $\rho$ increases with $|\Omega|$ and decreases with $N$.
For large $|\Omega|$, on the other hand, the average power deviates from ${p_0=0.5}$, as seen in Fig.~\ref{fig_stm_pc_n_cpl}(d), and thus nonlinear effects of $\delta\omega_i(p_i)$ and $\Gamma_i(p_i)$ in Eq.~(\ref{eq_nlo1}) can appear.
In this situation, it is known that the power degree of freedom causes disordered unsteady states, including chaotic states~\cite{Matthews1990, Acebron2005}.
In these states the synchronization is suppressed, which corresponds to low $\rho$ for ${|\Omega|\gtrsim0.15}$ ns$^{-1}$ in Fig.~\ref{fig_stm_pc_ord_cpl_n}(c).
These disordered unsteady states~\cite{Matthews1990, Acebron2005} can also explain the large standard deviations observed for the same region in Fig.~\ref{fig_stm_pc_ord_cpl_n}(c), which indicates that the oscillation states are sensitive to small changes in the conditions.
Between the synchronized state and the disordered state, for small and large $|\Omega|$, respectively, a transition occurs, causing a sharp change in $\rho$.

The decrease of $\rho$ with increasing $N$ is due to the short-range interaction between the STOs.
By designing the STOs in the array to couple through a long-range interaction, e.g., by using a common electrode~\cite{Kudo2017}, one can make $\rho$ finite for large $N$~\cite{Kuramoto1987, Acebron2005}, allowing the synchronization of a large number of STOs.

\section{Summary and outlook}
We investigate reservoir computing on an STO array coupled by dipolar fields.
Numerical experiments on reservoir computing are performed by using a corresponding nonlinear oscillator model, which exhibits synchronized oscillation.
As a result, we show that the performance of reservoir computing is improved by increasing the number of STOs in the synchronized oscillation state.
The values of the STM and PC capacities for ${N\leq400}$ can be larger than those for an ESN with 500 units.
In particular, the maximum STM capacity is more than twice that of an ESN.
The largest capacities are obtained at the boundary between the synchronized and disordered states.
We believe that the results can be explained by an enhancement of the capacities at the edge of chaos.
The present results indicate that STO arrays are promising for the hardware implementation of reservoir computing.
The performance can be further improved by introducing methods in reservoir computing not included here, such as time multiplexing.
In this paper, the effects of thermal fluctuations~\cite{Quinsat2010}, which are large in nanometer-sized magnetizations, are not included.
This issue is left for future work.

To realize the reservoir computing on an STO array, further considerations of several factors are necessary.
We expect that an array of pillar-shaped magnetic tunnel junctions (MTJs) like those in magnetoresistive random access memories~\cite{Bhatti2017} or an array of giant magnetoresistive elements can be utilized as the STO array.
MgO-based MTJs can exhibit large output signals due to the tunneling magnetoresistive effect.
Typical ferromagnetic materials based on elements such as Fe and Co will yield enough coupling by dipolar fields, as estimated in Appendixes~\ref{sec_sto_llgs} and \ref{sec_llgs_nlom}.
At the current time, a few STOs have been coupled by dipolar fields~\cite{Locatelli2015a}.
To increase the number of STOs, a one-dimensional array is the first candidate because the design of the electrodes which apply currents and extract signals will be simpler.
With an increasing number of lines, the array approaches a two-dimensional one.
Three-dimensional design of the electrodes will be necessary for integrating a large number of STOs into a two-dimensional array.
Instead, reservoir computing may be possible when the number of measured signals is smaller than the number of STOs, for example by measuring signals from a fraction of the STOs or measuring sums of signals from several STOs, which reduces the complexity of the electrode configuration.
In this case, estimation of the performance degradation caused by the reduced number of measured signals is necessary.
Although we have assumed that the input signals are applied to one STO, reservoir computing can be performed by applying input signals to several STOs, which increases the freedom of design of the array.
Input signals with waveforms other than a square wave, such as a distorted square wave or a triangle wave, will introduce similar fluctuations of the powers when the rates and amplitudes are suitable.
In the measurement of the powers as the outputs, signal-processing technologies for amplitude modulation can be used~\cite{Nakamura2018}.

\begin{acknowledgments}
The authors would like to thank S. Tsunegi, M. Goto, K. Kudo, and R. Sato for valuable discussions.
\end{acknowledgments}

\appendix
\section{SYNCHRONIZATION OF STOS IN LLGS EQUATION}\label{sec_sto_llgs}
We show an example of the magnetization dynamics of STOs by solving the LLGS equation.
The LLGS equation for the magnetizations of a free layer of STOs ${\bf m}_i$ ($i=1,2,\cdots,N$) normalized by the saturation magnetization $M_s$ is given by
\begin{eqnarray}
	\frac{\partial{\bf m}_i}{\partial t}=-\gamma{\bf m}_i\times{\bf H}^{\rm{eff}}_i+\alpha{\bf m}_i\times\frac{\partial{\bf m}_i}{\partial t}+{\bf T}^S_i,\label{eq_llgs}
\end{eqnarray}
where $N$, $t$, $\gamma$, and $\alpha$ are the number of STOs, the time, the gyromagnetic ratio, and the Gilbert damping, respectively.
${\bf H}^{\rm{eff}}_i$ and ${{\bf T}^S_i=\sigma I_i{\bf m}_i\times({\bf m}_i\times{\bf e}_p)}$ are an effective field and a spin-torque term with efficiency $\sigma$, current $I_i$, and unit vector in the direction of the spin polarization of the current ${\bf e}_p$.

\begin{figure}[t]
	\centering
	\includegraphics[width=8.5cm,bb=0 0 339 135]{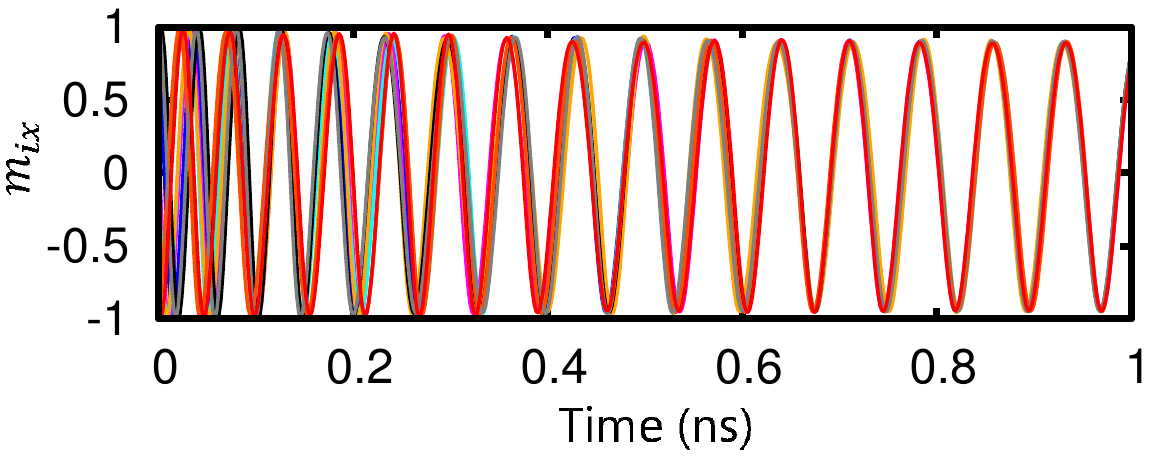}
	\caption{Waveforms of $m_{ix}$ for free layers of nine STOs obtained from the LLGS equation.
		${M_s=1.6\times10^3}$ emu/cm$^3$, ${\alpha=0.02}$, and ${H_z=9}$ kOe.
		Owing to $H_z$, ${{\bf e}_p=(\cos5^\circ,0,\sin5^\circ)}$ is assumed.
		Neighboring STOs are separated by $50$ nm.
		The mean and standard deviation of $I_i$ are $20$ mA and $0.2$ mA, respectively.
		The initial directions of ${\bf m}_i$ are randomized in the plane of ${m_{iz}=0}$.}
	\label{fig_mx_time}
\end{figure}

In the example, we choose an array of typical STOs with in-plane magnetized free and fixed layers, having a rectangular shape.
A small (${50^2\times10}$ nm$^3$) and thus uniform ${\bf m}_i$ is assumed.
A perpendicular field $H_z$ is applied, yielding an out-of-plane precession (OPP) for large $I_i$~\cite{Kiselev2004}.
The STOs are placed on a square lattice, and all the magnetizations ${\bf m}_i$ are coupled by dipolar fields~\cite{Kudo2012, Kudo2015}.
$I_i$ and the initial directions of ${\bf m}_i$ are randomized.
Figure~\ref{fig_mx_time} shows an example of the waveforms.
Owing to the randomness, the waveforms are initially disordered.
After several oscillation periods, the oscillations are synchronized.
This state is obtained for other initial conditions, indicating the robustness of the synchronized state.
This kind of synchronization has been reported in vortex-type STOs coupled by dipolar fields~\cite{Belanovsky2012, Locatelli2015a, Flovik2016}.

\section{FROM THE LLGS EQUATION TO THE NONLINEAR OSCILLATOR MODEL}\label{sec_llgs_nlom}
\subsection{Coupling between STOs}\label{sec_coupling}
Here, we show the correspondence between the LLGS equation and the nonlinear oscillator model.
In the OPP, the $x$ and $y$ components of ${\bf m}_i$ are mainly oscillating, and we introduce complex amplitudes ${\tilde c_i=m_{ix}-\mathrm{i}m_{iy}}$.
From Eq.~(\ref{eq_llgs}), the equation of motion of $\tilde c_i$ is
\begin{eqnarray}
	\frac{\mathrm{d}\tilde c_i}{\mathrm{d} t}=-\mathrm{i}\gamma H^{\rm eff}_{iz}\tilde c_i+\mathrm{i}\gamma m_{iz}\left(H^{\rm eff}_{ix}-\mathrm{i}H^{\rm eff}_{iy}\right)-\Gamma_i\tilde c_i,\label{eq_nlo}
\end{eqnarray}
where the term including ${\bf H}^{\rm{eff}}_i$ is transformed exactly, while the damping and spin-torque terms are represented by using an effective damping rate $\Gamma_i$, which is discussed in Sec.~\ref{sec_EffectiveDamping}.
Here, we assume uniform magnetizations in each STO.
In the OPP, $H^{\rm eff}_{iz}$ mainly determines the precession axis and frequency ${\tilde\omega_i=\gamma H^{\rm eff}_{iz}}$.
Besides, ${H^{\rm eff}_{ix}-\mathrm{i}H^{\rm eff}_{iy}}$ is due to the dipolar fields from the other STOs.
Although the prefactor includes $m_{iz}$, this prefactor depends little on $\tilde c_i$, because $m_{iz}$ is almost constant for the OPP.
The term including ${H^{\rm eff}_{ix}-\mathrm{i}H^{\rm eff}_{iy}}$ reduces to the coupling term as follows.
We approximate, for simplicity, the magnetization of a free layer of volume $V$ by a point magnetic moment ${{\bm\mu}=M_sV{\bf m}}$.
The dipolar field at the position ${{\bf r}=(x, y, z)}$ arising from this moment at the origin is given by
\begin{eqnarray}
{\bf H}=3\frac{\left(\bm{\mu}\cdot{\bf r}\right){\bf r}}{r^5}-\frac{{\bm\mu}}{r^3},
\end{eqnarray}
Then,
\begin{eqnarray}
	&&H_x-\mathrm{i}H_y=\frac{M_sV}{2r^5}\\
	&&\times\left[\left(x^2+y^2-2z^2\right)\tilde c+3(x-\mathrm{i}y)^2\tilde c^*+6(x-\mathrm{i}y)zm_z\right].\nonumber
\end{eqnarray}
On the same plane, i.e., ${{\bf r}=(x,y,0)}$,
\begin{eqnarray}
	H_x-\mathrm{i}H_y=\frac{M_sV}{2r^3}\left[\tilde c+3\frac{(x-\mathrm{i}y)^2}{r^2}\tilde c^*\right].
\end{eqnarray}
Thus Eq.~(\ref{eq_nlo}) can be written as
\begin{eqnarray}
	\frac{\mathrm{d}\tilde c_i}{\mathrm{d}t}=-\mathrm{i}\tilde\omega_i\tilde c_i-\mathrm{i}\sum^N_{j=1,j\neq i}\left(\Omega_{ij}\tilde c_j+\Xi_{ij}\tilde c^*_j\right)-\Gamma_i\tilde c_i,
\end{eqnarray}
where ${\Omega_{ij}=-\gamma m_{iz}M_sV/\left(2|{\bf r}_i-{\bf r}_j|^3\right)}$ and $\Xi_{ij}=-3\gamma m_{iz}M_sV\left[x_i-x_j-\mathrm{i}\left(y_i-y_j\right)\right]^2/\left(2|{\bf r}_i-{\bf r}_j|^5\right)$.
Here ${\bf r}_i$ represents the position of STO $i$.

We assume that the STOs generate a common frequency $\omega$ when isolated in the steady state, and that the variation of the frequencies is caused by the nonlinear frequency shifts $\delta\omega_i$.
Thus, the oscillation frequencies are represented by ${\tilde\omega_i=\omega+\delta\omega_i}$.
Then, the uniform oscillation with frequency $\omega$ is decoupled by ${\tilde c_i=c_i\mathrm{e}^{-\mathrm{i}\omega t}}$ as
\begin{eqnarray}
	\frac{\mathrm{d}c_i}{\mathrm{d}t}=-\mathrm{i}\delta\omega_ic_i-\mathrm{i}\sum_j\left(\Omega_{ij}c_j+\Xi_{ij}c^*_j\mathrm{e}^{2\mathrm{i}\omega t}\right)-\Gamma_ic_i.\hspace{1em}\label{eq_nlo2}
\end{eqnarray}
Note that an oscillating factor with a frequency $2\omega$ appears in the coefficient of $c^*_j$.
This term does not affect the dynamics of $c_i$ when $c_i$ varies slowly, as shown in the following.
We assume that the timescale of the change of $c_i$ is much longer than the oscillation period ${2\pi/\omega}$ and that $c_i$ can be approximated by a constant during $2\pi/\omega$.
Then, by averaging Eq.~(\ref{eq_nlo2}) for a time $\pi/\omega$, the term including $c^*_j$ vanishes owing to the factor $\mathrm{e}^{2\mathrm{i}\omega t}$, and hence the nonlinear oscillator model,
\begin{eqnarray}
	\frac{\mathrm{d}c_i}{\mathrm{d}t}=-\mathrm{i}\delta\omega_ic_i-\Gamma_ic_i-\mathrm{i}\sum^N_{j=1,j\neq i}\Omega_{ij}c_j,
\end{eqnarray}
is obtained.
This averaging is equivalent to using a variable
\begin{eqnarray}
	\bar c_i(t)=\frac{\omega}{\pi}\int^{\pi/\omega}_0\mathrm{d}t'c_i(t+t').
\end{eqnarray}

The numerical order of magnitude of $\left|\Omega_{ij}\right|$ can be estimated as follows.
In the typical case of ${m^2_{iz}=0.5}$, ${M_s=1.6\times10^3}$ emu/cm$^3$, and ${V=50^2\times10}$ nm$^3$, $\left|\Omega_{ij}\right|$ is $0.031$ ns$^{-1}$ and $0.25$ ns$^{-1}$ for center-to-center distances of $200$ nm and $100$ nm, respectively, indicating that $\left|\Omega_{ij}\right|$ is of the order of $0.01$ ns$^{-1}$ to $0.1$ ns$^{-1}$.

\subsection{Effective damping rate and nonlinear frequency shift}\label{sec_EffectiveDamping}
The effective damping rate $\Gamma_i$ is assumed to be a function of the current $I_i$, while the nonlinear frequency shift $\delta\omega_i$ is not~\cite{Slavin2009}.
The reason for this can be illustrated as follows.
Since $\Gamma_i$ and $\delta\omega_i$ describe the properties of a single STO, we focus on the single-STO case, in which the LLGS equation~(\ref{eq_llgs}) is written as
\begin{eqnarray}
	\frac{\partial{\bf m}}{\partial t}&\simeq&-\gamma{\bf m}\times{\bf H}^{\rm{eff}}-\alpha\gamma{\bf m}\times\left({\bf m}\times{\bf H}^{\rm{eff}}\right)\nonumber\\
	&&+\sigma I{\bf m}\times\left({\bf m}\times{\bf e}_p\right).\label{eq_llgs_sin}
\end{eqnarray}
Here, the Gilbert damping term is transformed to this form [the second term in the right-hand side of Eq.~(\ref{eq_llgs_sin})] by assuming ${\alpha^2\ll1}$ and ${|\alpha\sigma I|\ll1}$, which is valid for typical STOs.
We consider the OPP and assume ${\bf H}^{\rm{eff}}$ in the $z$ direction.
By supposing ${{\bf e}_p=e_{px}\hat{{\bf x}}+e_{pz}\hat{{\bf z}}}$ (where $\hat{{\bf x}}$ and $\hat{{\bf z}}$ are unit vectors in the $x$ and $z$ directions), the spin-torque term can be divided into two terms.
The term including $e_{pz}\hat{{\bf z}}$ is collinear with the Gilbert damping term and can balance and cancel that term, allowing a stationary oscillation state.
From these terms, the effective damping rate can be expressed as ${\Gamma=\left(\alpha\gamma H^{\rm{eff}}_z-\sigma Ie_{pz}\right)m_z}$, which depends explicitly on $I$ and is almost constant for the OPP.
On the other hand, the term including $e_{px}\hat{{\bf x}}$ can have a component parallel to the first torque term ${-\gamma{\bf m}\times{\bf H}^{\rm{eff}}}$, which determines the oscillation frequency.
The total contribution of the $e_{px}\hat{{\bf x}}$ term during one precession, however, is small because the component in the direction of the first torque term changes sign during the precession and cancels out~\cite{Taniguchi2013}.
The component in the direction of the first torque is proportional to ${-\left({\bf m}\times{\bf H}^{\rm{eff}}\right)\cdot\left[{\bf m}\times\left({\bf m}\times e_{px}\hat{{\bf x}}\right)\right]=H^{\rm{eff}}_ze_{px}m_y}$, and the sign of $m_y$ changes for the OPP during one precession.
Thus the dependence of the oscillation frequency on the current can be neglected, and it is assumed that $\delta\omega$ is not a explicit function of $I$.
Note that $\delta\omega$ is indirectly affected by $I$ as a result of changes in the power $p$.

The forms of $\Gamma_i$ and $\delta\omega_i$ used in Eqs.~(\ref{eq_omega}) and (\ref{eq_gamma}) are obtained as follows.
In a stationary oscillation state of an individual STO with power $p_{i0}$ and current $I_{i0}$, $\Gamma_i$ is zero because the spin-transfer torque and the Gilbert damping are balanced and cancel each other.
When a small modulation of $I_i$ is introduced as an input signal, $p_i$ deviates from $p_{i0}$ as a result of time evolution.
Then $\Gamma_i$ can be linearly approximated as
\begin{eqnarray}
	\Gamma_i&\simeq&\frac{\partial\Gamma_i}{\partial p_i}\left(p_i-p_{i0}\right)+\frac{\partial\Gamma_i}{\partial I_i}\left(I_i-I_{i0}\right)\\
	&=&\frac{\Gamma_p}{p_{i0}}\left(p_i-p_{i0}\right)-\xi_i(t),
\end{eqnarray}
where the first term is expressed by using a damping rate of the power~\cite{Slavin2009}
\begin{eqnarray}
	\Gamma_p=p_{i0}\frac{\partial\Gamma_i}{\partial p_i}.
\end{eqnarray}
We express the second term yielded by the input signal as
\begin{eqnarray}
	\frac{\partial\Gamma_i}{\partial I_i}\left(I_i-I_{i0}\right)=-\xi_i(t).
\end{eqnarray}
The minus sign is because the spin-transfer torque works as a negative damping, which is proportional to the current.
As we assume that ${\delta\omega_i=0}$ at ${p_i=p_{i0}}$ and $\delta\omega_i$ does not depend explicitly on $I_i$, $\delta\omega_i$ is approximated as ${\delta\omega_i\simeq N_f\left(p_i-p_{i0}\right)}$.
This kind of description of the current modulation has been developed in Ref.~\cite{Slavin2009}.

\section{NARMA TASK}\label{sec_narma}
Here, we show the performance in the NARMA task, where series defined by nonlinear recurrence relations are emulated, and the ability to predict nonlinear time series is measured.
We perform a NARMA task defined by the following equation: $y_{k+1}=0.4y_k+0.4y_ky_{k-1}+0.6s^3_k+0.1$ for $s_k$ consisting of uniform random numbers in ${(0,0.2)}$, called the NARMA2 task~\cite{Fujii2017, Atiya2000}.
We convert $s_k$ to $\xi_1(t)$ by use of a square wave with value ${\xi_1(t)=s_k-0.1}$ ns$^{-1}$.
The performance is quantified by the normalized mean-squared error
${\epsilon=\left[\sum_k\left(\bar{y}_k-y_k\right)^2\right]/\left(\sum_k\bar{y}^2_k\right)}$, which is smaller when $\bar{y}_k$ and $y_k$ are closer, indicating higher performance.

\begin{figure}[h]
	\centering
	\includegraphics[width=8.5cm,bb=0 0 225 147]{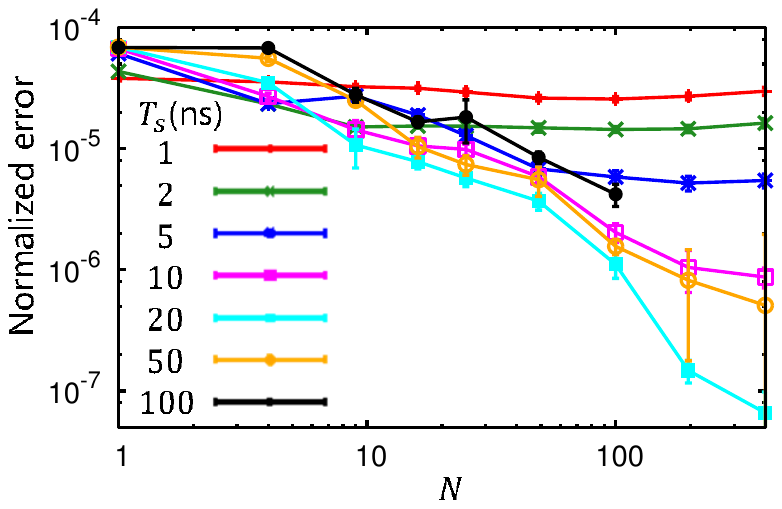}
	\caption{Normalized mean-squared error $\epsilon$ in NARMA2 task as a function of $N$ and $T_s$.}
\label{fig_nar2}
\end{figure}

The results for the NARMA2 task are shown in Fig.~\ref{fig_nar2}.
The error for ${T_s=20}$ ns decreases with increasing $N$ by three orders of magnitude, and the value of the error is comparable to that in previous studies~\cite{Fujii2017}.

\end{document}